\documentclass[12pt]{article}
\usepackage{graphicx}
\usepackage{amssymb}
\usepackage{amscd}
\usepackage{amsmath}
\usepackage{appendix}
\usepackage{bbm}
\usepackage[table]{xcolor}
\usepackage{multirow}
\usepackage{array}
\usepackage[normalem]{ulem}
\usepackage{tabularx,booktabs}

\begin{document}
\begin{titlepage}

{\hbox to\hsize{\hfill December 2015 }}

\bigskip \vspace{3\baselineskip}

\begin{center}
{\bf \large 

Electroweak Baryogenesis with Anomalous Higgs Couplings }

\bigskip

\bigskip

{\bf  Archil Kobakhidze, Lei Wu and Jason Yue \\ }

\smallskip

{ \small \it
ARC Centre of Excellence for Particle Physics at the Terascale, \\
School of Physics, The University of Sydney, NSW 2006, Australia \\
E-mails: archil.kobakhidze, lei.wu1, jason.yue@sydney.edu.au}

\bigskip
 
\bigskip

\bigskip

{\large \bf Abstract}

\end{center}
\noindent 
We investigate feasibility of efficient baryogenesis at the electroweak scale within the effective field theory framework based on a non-linear realisation of the electroweak gauge symmetry. In this framework the LHC Higgs boson is described by a singlet scalar field, which, therefore, admits new interactions. Assuming that Higgs couplings with the eletroweak gauge bosons are as in the Standard Model, we demonstrate that the Higgs cubic coupling and the CP-violating Higgs-top quark anomalous couplings alone may drive the a strongly first-order phase transition. The distinguished feature of this transition is that the anomalous Higgs vacuum expectation value is generally non-zero in both phases. We identify a range of anomalous couplings, consistent with current experimental data, where sphaleron rates are sufficiently fast in the 'symmetric' phase and are suppressed in the 'broken' phase and demonstrate  that the desired baryon asymmetry can indeed be generated in this framework. This range of the Higgs anomalous couplings can be further constrained from the LHC Run 2 data and be probed at high luminosity LHC and beyond.              
 
\end{titlepage}


\numberwithin{equation}{section}
\renewcommand{\theequation}{\thesection.\arabic{equation}} 

\section{Introduction}
The discovery of a 125 GeV Higgs particle at LHC \cite{Aad:2012tfa, Chatrchyan:2012xdj} has important implications for cosmology. Namely, with the Higgs data available, we can now analyse the nature of the electroweak phase transition and baryogenesis \cite{Kuzmin:1985mm}-\cite{Nelson:1991ab} in a more quantitative manner. It has been established already well before the Higgs discovery that within the Standard Model the required first-order phase transition is not realised for the Higgs boson heavier than $\sim 70$ GeV \cite{Kajantie:1996mn}  and the standard Cabibbo-Kobayashi-Maskawa CP-violation is not sufficient to generate the desired baryon asymmetry \cite{Gavela:1994dt, Huet:1994jb}. Hence, the observed matter-antimatter asymmetry \cite{Cooke:2013cba, Ade:2015xua}:
\begin{equation}
\eta_B:= n_B/s \sim8.6 \times10^{-11}~,
\label{1}
\end{equation} 
[here $n_B$ and $s$ denote respectively the net baryon number and entropy densities] provides with a strong hint in favour of new physics beyond the Standard Model. For an overview and current status of the electroweak baryogenesis readers are referred to the reviews \cite{Cohen:1993nk, Trodden:1998ym, Morrissey:2012db} and references therein.

Such new physics can be encoded in an effective field theory (EFT) extension of the Standard Model which contains a set of nonrenormalizable, gauge invariant operators of canonical dimension greater than 4.  The lowest dimensional nonrenormalizable operators, which are relevant  for the electroweak baryogenesis \cite{Zhang:1993vh}-\cite{Huang:2015bta} and are relatively less constrained at present \cite{Ellis:2014jta} are:
\begin{equation}
\mathcal{O}_{6} = \frac{c_{6}}{\Lambda^2}(H^\dagger H)^3 , \quad \mathcal{O}_{t\overline{t} h}  = \frac{c_{t\bar th }}{\Lambda^2}(H^\dagger H) \overline{Q}_L H t_R~,
\label{2}  
  \end{equation}
where $\Lambda$ is a cut-off scale and $c_{6}$ and $c_{t\bar th }$ are the dimensionless constants $\sim {\cal O}(1)$. For the Higgs boson of mass $m_h=125$ GeV the successful baryogenesis requires the low ultraviolet cut-off, $\Lambda 
\lesssim 840$ GeV or so \cite{Grojean:2004xa}. This shades some doubts on the validity of the EFT framework. Several specific renormalizable models of the electroweak baryogenesis in light of the LHC Higgs data have been also discussed recently in \cite{Chung:2012vg}-\cite{Damgaard:2015con}.      

Although the current Higgs data is consistent with the Standard Model predictions, one should bear in mind that the nature of the electroweak symmetry breaking is not fully understood yet. In particular, instead of the standard linear realisation the electroweak gauge symmetry can be nonlinearly realised with the Higgs boson residing in the singlet representation of the symmetry group \cite{Kobakhidze:2012wb} (see also \cite{Ferrara:1992sd, Kobakhidze:2015loa} for supersymmetric models). Nonlinearly realised electroweak gauge theory becomes strongly interacting at high energies, the famous example being $WW\to WW$ scattering in the Higgsless Standard Model (SM). It is expected that at high energies new resonances show up, which unitarise rapid, power-law  growth of scattering amplitudes with energy in perturbation theory. However, the scale where new physics is expected to emerge crucially depends on the specific process considered. For example, in SM with the anomalous top-Yukawa couplings perturbative unitarity is violated for the $t\bar t \to WW$ process at energies $\sim 10$ TeV \cite{Appelquist:1987cf}. New physics at such high energies may escape the detection at LHC. In situations like these, precision measurements of deviations from SM physics parametrized within the effective theories based on nonlinear realisation become imperative.  

The purpose of this work is to investigate feasibility of efficient baryogenesis at the electroweak scale within the EFT framework based on a non-linear realisation of the electroweak gauge symmetry. Namely, we consider a model with a modified Higgs-Yukawa sector only \cite{Kobakhidze:2012wb}. It admits a tree-level cubic Higgs coupling and CP-violating Higgs-Top interactions, which play a significant role in the electroweak phase transition and baryogenesis. This model is described in the next section. In section 3 we compute a finite temperature effective potential and analyse the phase transition. A remarkable consequence of the model with anomalous tree-level cubic Higgs and Higgs-Top couplings is that the Higgs vacuum expectation value does not vanish at high temperatures. Taking into account current constraints on the anomalous Higgs-Top couplings \cite{Kobakhidze:2014gqa}, we identify a parameter range where the electroweak phase transition is strongly first-order. We then proceed with computing the baryon asymmetry in section 4. We conclude in section 5. Some technicalities are collected in the Appendices for the reader's convenience.        

\section{Description of the model}
The electroweak symmetry within SM is spontaneously broken via nonzero vacuum expectation value $v\approx 246$ GeV of the $SU(2)\times U(1)_Y$ complex doublet Higgs field $H(x)=\left(H^{+}(x), H^0(x)\right)^{\rm T}$, where one neutral scalar component is the physical Higgs field, while the remaining three components describe longitudinal degrees of freedom of $W^{\pm}, Z^0$ vector bosons. In our model, the later degrees of freedom are combined  in nonlinear field ${\cal X}(x)$:
\begin{equation}
	{\cal X}(x)=e^{\frac{i}{2}\pi_i(x)T_i}\left(
\begin{tabular}{c}
	0 \\ 
	1 \\ 
	\end{tabular} 	
	\right)
	\label{3}
\end{equation}  
where $T_{i}=\sigma_{i}-\mathbbm{1}\delta_{i3}$ are the three broken generators  for $i\in\{1,2,3\}$,  with  $\sigma_i$ denoting the Pauli matrices and  $\pi_i (x)$ are the three would-be Goldstone bosons spanning $SU(2)\times U(1)_Y/U(1)_{EM}$ coset space. 
With non-linear realization of $SU(2)\times U(1)_Y$ electroweak gauge invariance the Higgs field $h$ is no longer obliged to form the electroweak doublet irreducible representation. We entertain the possibility that the Higgs boson resides in $SU(2)\times U(1)_Y$ singlet field $\rho(x)$. The standard Higgs doublet then can be identified with the following composite field\footnote{We note that if $\rho(x)$ field is to be identified with the modulus of the electroweak doublet field, $\rho^2=H^{\dagger}H$, it should be restricted to positive ($\rho>0$) or negative ($\rho<0$) values only.}:
\begin{equation}
H(x)=\frac{\rho(x)}{\sqrt{2}}{\cal X}(x)~.
\label{4}
\end{equation}

While maintaining $SU(2)\times U(1)_Y$ invariance, the non-linear realisation of the electroweak gauge symmetry allows a number of new interactions beyond those present in SM. A generic model is severely constrained by the electroweak precision measurements, flavour physics and the Higgs data. Therefore, here we consider only extra interactions which are most relevant for the electroweak baryogenesis and are relatively less constrained by current data. Namely, we consider CP-violating Higgs-top Yukawa interactions, which, in the basis of diagonal up-quark Yukawa matrix, looks as: 
\begin{eqnarray}
{\cal L}_{\rm Higgs-Top}=-\left[m'_{t}+Y_{t}\rho/\sqrt{2}\right]\bar Q_L\tilde{{\cal X}}t_{\rm R}+{\rm h.c.}
~,  \label{5}
\end{eqnarray}
where $Q_L=(t_L, b_L)^{\rm T}$ is the third generation left-handed quark doublet, $\tilde{{\cal X}}=i\sigma^2{\cal X}^{*}$, $m'_t$ is an additional mass parameter and $Y_{t}=y_t{\rm e}^{i\xi}$ is a complex coupling.  

In addition to Eq. (\ref{5}) we consider modified Higgs potential, which contains tree-level cubic term: 
\begin{equation}
V(\rho)=-\frac{\mu^2}{2}\rho^2+\frac{\kappa}{3} \rho^3+\frac{\lambda}{4}\rho^4~,
\label{6}
\end{equation}      
We assume that the scalar potential has a global minimum for a non-zero vacuum expectation value of the Higgs field $\rho$ (see the next section): 
\begin{equation}
\langle \rho\rangle = v~,~~|v|\approx 246~{\rm GeV}~,
\label{7}
\end{equation}
The absolute value of the vacuum expectation value in (\ref{7}) is fixed to the standard value since the Higgs interactions with the electroweak gauge bosons are assumed to be the same as in SM, i.e.,
\begin{equation}
\frac{\rho^2}{2}(D_{\mu}{\cal X})^{\dagger}D^{\mu}{\cal X}~,
\label{8}
\end{equation}
where $D_{\mu}$ is an $SU(2)\times U(1)_Y$ covariant derivative. The shifted field
\begin{equation}
h(x)=\rho(x) - v
\label{9}
\end{equation}
describes the physical excitation associated with the Higgs particle with the tree-level mass squared:
\begin{equation}
m_h^2=\left.\frac{\partial^ 2 V}{\partial \rho\partial \rho}\right\vert_{\rho=v}\approx \left( 125~{\rm GeV}\right)^2~.
\label{10}
\end{equation} 

The tree-level top quark mass $m_t\approx 173$ GeV is defined  in our model from Eq. (\ref{5}) upon expressing $\rho$ in terms of $h$ and $v$ using Eq. (\ref{9}). It is then convenient to express the extra mass parameter  $m'_t$ through  $\xi$ and $y_t$, and $m_t$. Due to the quadratic relation we account an ambiguity in definition of $m'_t$:
 \begin{equation}
{m'_t}^{(\pm)}=\frac{1}{2}  \left(\pm\sqrt{ 4m_t^2 - 2y_t^2 v^2 \sin^2\xi }- \sqrt{2}y_t v \cos \xi \right)~.
\label{11}
\end{equation}

Note that, the cubic interaction term in (\ref{2}) explicitly breaks the discrete $\rho\to -\rho$ symmetry, thus avoiding potentially dangerous cosmological domain wall problem. More importantly, as will be shown below, it plays a crucial role in enhancing the first-order electroweak phase transition and, together with the additional  additional CP-violation appearing in Higgs-top Yukawa interactions, leads to the the successful electroweak baryogenesis. 

Remarkably, the electroweak precision observables parametrized through the $S,T,U$ oblique parameters are essentially unaffected in our model. This is because we have not introduced any new particle and the 1-loop oblique parameters depend on particle masses rather than Higgs-Yukawa couplings. Also, the current data is not sensitive to the Higgs self-interaction couplings and they are unconstrained at present. In addition, the above model satisfies the weak power counting renormalisability as discussed in \cite{Binosi:2012cz}. 

The Higgs data constraints on CP-violating Higgs-top couplings have been discussed in \cite{Kobakhidze:2014gqa}. We take these constraints into account in what follows. Finally, we note that the new source of CP-violation in the Higgs-top sector induces additional contribution to the electric dipole moments (EDMs) of charged fermions, $d_{f}$. Following \cite{Barr:1990vd}, we have computed this contribution  in our model:
\begin{eqnarray}
d_f/\bar d_f=\frac{4}{9}\sin{\xi}\left(\frac{m_t'}{m_t}\right)\left(\frac{y_tv}{\sqrt{2}m_t}\right)\ln\left(\frac{m_t^2}{m_h^2}\right) \nonumber \\
\approx 0.29\sin \xi \left(\frac{y_t}{y_t^{\rm SM}}\right)\left(\frac{m_t'}{m_t}\right)~.
\end{eqnarray}
 where $\bar d_f=\frac{|Q_f|\alpha m_f}{16 \pi^3v^2}$, e.g., $\bar d_e \approx 2.5\cdot 10^{-27}$ e$\cdot$cm, and $y_t^{\rm SM}$ is the SM top-Yukawa coupling.  We found that the most of the parameter space allowed by the Higgs data is also consistent with the experimental constraints on the electron EDM, 
 $|d_e| < 8.7\times 10^{-29}$, established at 90\% confidence level using polar thorium monoxide (THO) molecules \cite{Baron:2013eja} (see also \cite{Chien:2015xha}). One should also bear in mind that possible anomalous couplings with other SM fermions may lead accidental cancellations in the fermion EDMs.  

\section{The electroweak phase transition}
The presence of the cubic term in the tree-level Higgs potential (\ref{6}) and the anomalous Higgs-top Yukawa couplings (\ref{5}) significantly alter the Higgs vacuum configuration. In this section we discuss the Higgs vacuum at zero and finite temperatures and the corresponding electroweak phase transition.  
 
\subsection{Higgs vacuum at zero temperature}
We find convenient to rewrite the mass parameter $\mu^2$ and the quartic coupling $\lambda$ in terms of the (tree level) Higgs mass, $m_h\approx 125$ GeV, the Higgs vacuum expectation value $v$ (\ref{7}) and the cubic coupling $\kappa$:
 \begin{eqnarray}
\mu^2& =\frac{1}{2}\left( m_h^2+v\kappa\right), \label{12a}  \\
\lambda &=\frac{1}{2v^2}\left( m_h^2-v\kappa\right). 
\label{12}
\end{eqnarray}   
The potentially must be bounded from below, that is $\lambda >0$ and, hence, $v\kappa < m_h^2$. There are three cases to consider:
\begin{itemize}
\item[i.] The non-tachyonic mass parameter, i.e., $\mu^2<0$ or, equivalently, $v\kappa < -m_h^2$.  One of the local minima in this case is at a trivial configuration $\langle \rho \rangle=0$. We find that the electroweak symmetry breaking minimum (\ref{7}) is realised as an absolute minimum of the potential if $-3m_h^2<v \kappa <-m_h^2$;
\item[ii.] The tachyonic mass parameter, i.e., $\mu^2>0$, which, in turn, implies $v\kappa > -m_h^2$.  In this case the trivial configuration is a local maximum and the minimum (\ref{7}) is realised providing $-m_h^2<v\kappa <0$. 
\item[iii.] For $\mu^2=0$ ($v\kappa =-m_h^2$), $v=-\frac{\kappa}{\lambda}$. In this case there are two trivial solutions for the extremum equation, which represent an inflection point of the potential.  
\end{itemize}
Notice, the symmetry of the above vacuum solutions under $\kappa \to - \kappa$ and $v\to -v$. 

Within the perturbation theory, one expects that the above tree-level analysis modifies insignificantly, except the case when the tree level cubic parameter is vanishingly small, $\kappa \approx 0$. In this case the radiative corrections induced by the anomalous Higgs-top Yukawa interactions (\ref{5}) must be taken into account. At one loop level this can be achieved by the substitution:
\begin{equation}
\kappa \to \kappa -\frac{3\sqrt{2}m'_ty_t^3\cos\xi}{16\pi^2}\left[2\cos^2 \xi+3\ln\left(\frac{m_t^{\prime 2}}{v^2}\right)\right]~.
\label{13}
\end{equation}   

\begin{figure}[t]
\begin{center}
\includegraphics[width=0.7\textwidth]{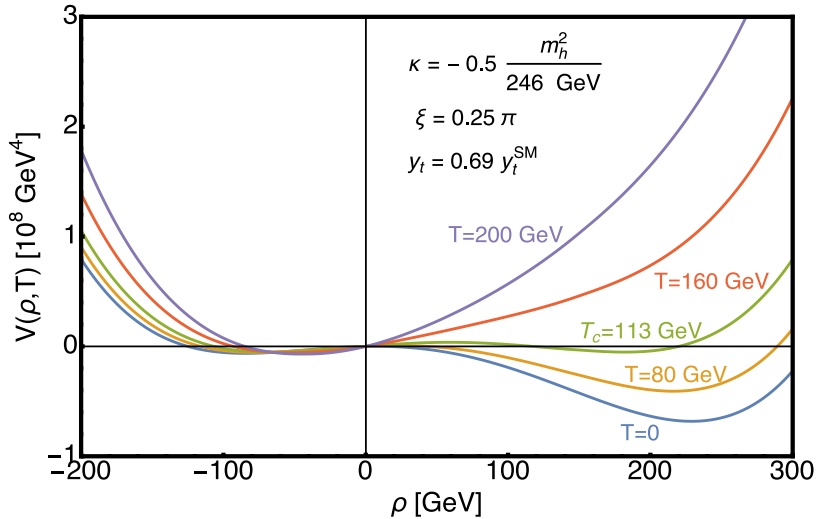}
\end{center}
\caption{{\small  Thermal effective potential at various temperatures. Non-zero thermal ground state $v^s_T$ persists at high temperatures. }}
\label{fig1}
\end{figure}


\subsection{Higgs vacuum at finite temperature}
The anomalous couplings have even more profound effect on the phase transition at finite temperature. To illustrate the main point let us first consider 1-loop finite temperature Higgs effective potential where only leading $T^2$-dependence on temperature is kept:
\begin{eqnarray}
V_T(\rho, T)&=&V(\rho)+\delta V_{\rm T}^{(1)} \nonumber \\
&\approx &V(\rho)+ \left(3g_2^2+g_1^2+4\lambda +4y_t^2\right)\frac{T^2\rho^2}{2}+\left(\kappa + 3\sqrt{2}y_tm'_t\cos\xi\right)\frac{T^2\rho}{12}~,
\label{14}
\end{eqnarray} 
where $g_{2,1}$ are $SU(2)$ and $U(1)_Y$ electroweak gauge couplings. Note that besides the standard $\sim T^2\rho^2$ thermal correction to the thermal Higgs potential, we account an additional term in Eq. (\ref{14}) which is linear in field $\rho$. This means that the gauge symmetry is never restored in our framework. Indeed, even for very large temperatures $T\gg\rho$, we will have non-zero expectation value,
\begin{equation}
v_T\approx - \frac{4}{3} \frac{\kappa+3\sqrt{2}y_t{m'_t}\cos \xi }{3{g^2_2}+  g^2_1 +4{\lambda} + 4y_t^2}~,
\label{14a}
\end{equation}   
which is proportional to the anomalous couplings and is essentially independent of $T$. This is ilustrated in Figure \ref{fig1} where the thermal effective potential is plotted for various temperatures. 

The above observation has an important implication for the baryogenesis scenario. On the one hand, we require strongly first-order phase transition with sphaleron effects \cite{Manton:1983nd, Klinkhamer:1984di} to be suppressed in the broken phase, i.e. $v^{br}_{T_c}/T_c > 1$. On the other hand, sphalerons must be effective in the 'symmetric' phase, i.e. $v^{s}_{T_c}/T_c<1$. Here $T_c$ denotes a critical temperature of the phase transition defined as: $v^s_{T_c}=v^{br}_{T_c}$. 

We have analysed the full one-loop thermal effective potential (see Appendix \ref{A}) and find a parameter area where the above requirements for the phase transition are satisfied. Some values of the critical temperature $T_c$ and thermal vacuum expectation values, $v^s_{T_c}$ and $v^{br}_{T_c}$ for various anomalous couplings $\kappa$, $y_t$ and $\xi$ are given in Table \ref{tab1}. We constraint the variation of $y_t$ and $\xi$ according to bounds obtained in our previous work \cite{Kobakhidze:2014gqa}. From these tables one can draw a picture of the phase transition in different areas of parameter space. Namely, for small $|\kappa|\lesssim 6$ GeV, the phase transition is not much different from the one in the Standard Model. The anomalous Higgs-top couplings are not large enough either to drive the first-order phase transition through the radiative contribution [see Eq. (\ref{13})]. As the magnitude of $\kappa$ increases, we observe several values of $v^s_{T_c}$ and $v^{br}_{T_c}$ that correspond to a strong first-order phase transition. The critical temperature $T_c$ decreases with the increase of $|\kappa|$ and reaches values as low as $\sim 50$ GeV. At the same time, $|v^s_{T_c}|$ decreases too. For very large $|\kappa|\gtrsim 130-160$ GeV, $v^s_{T_c}$ approaches to zero and essentially becomes an inflection point of the effective potential. Hence, the desired first-order phase transition ceases to exist for such large $|\kappa|$.                     

\begin{table}[hpt]
\fontsize{8pt}{9.6pt}\selectfont
\begin{center}
{\renewcommand{\arraystretch}{1.2}
\begin{tabular}{ c|c| ccc }
\hline 
{\multirow{2}{*}{$\kappa |v| /m_h^2$}}&&\multicolumn{3}{c}{$\xi=0$}\\ \cline{3-5}
&& ${y_t}=0.77{y_t^{SM}}$& ${y_t}=0.99{y_t^{SM}}$ &${y_t}=1.2{y_t^{SM}}$  \\\hline\hline
\multirow{2}{*}{$-0.1$}&$v^s_{T_c},~v^{br}_{T_c}$&$-189.$, $214$ & $-$ & $-$\\
&$T_c$&$77.6$ & $-$ &$-$\\\hline

\multirow{2}{*}{$-0.5$}&$v^s_{T_c},~v^{br}_{T_c}$&$-61.5$, $175.$ & $-$ & $-$\\
&$T_c$&$110.4$ & $-$ & $-$\\\hline

\multirow{2}{*}{$-1.0$}&$v^s_{T_c},~v^{br}_{T_c}$&$-19.9$, $187.$ & $20.8$, $144.$ & $-$\\
&$T_c$&$98.4$ & $104.$ & $-$\\\hline

\multirow{2}{*}{$-1.5$}&$v^s_{T_c},~v^{br}_{T_c}$&$-9.44$, $206.$& $2.90$, $190.$ & $41.7$, $151.$\\
&$T_c$&$80.5$ & $83.2$ & $92.1$\\\hline

\multirow{2}{*}{$-2.0$}&$v^s_{T_c},~v^{br}_{T_c}$&$-5.46$, $220.$& $-1.51$,  $211.$ & $6.82$, $200.$\\
&$T_c$&$53.0$ & $55.9$ & $63.4$\\\hline

\multirow{2}{*}{$-2.5$}&$v^s_{T_c},~v^{br}_{T_c}$&$-$& $-$ & $-$\\
&$T_c$&$-$ & $-$ & $-$\\\hline
\end{tabular}}
\end{center}

\vspace{-0.6cm}


\begin{center}
{\renewcommand{\arraystretch}{1.2}
\begin{tabular}{ c|c| ccc }
\hline 
{\multirow{2}{*}{$\kappa |v| /m_h^2$}}&&\multicolumn{3}{c}{$|\xi|=0.25\pi$}\\ \cline{3-5}
&& ${y_t}=0.62{y_t^{SM}}$& ${y_t}=0.69{y_t^{SM}}$ &${y_t}=0.76{y_t^{SM}}$  \\\hline\hline
\multirow{2}{*}{$-0.1$}&$v^s_{T_c},~v^{br}_{T_c}$&$-197.$, $224$&$-194.$, $219.$& $-187.$, $213.$\\
&$T_c$&$72.9$&$75.7$& $81.2$\\\hline

\multirow{2}{*}{$-0.5$}&$v^s_{T_c},~v^{br}_{T_c}$&$-73.4$, $188.$&$-67.5$, $182.$& $-58.8$, $174.$\\
&$T_c$&$114.$&$113.$& $113.$\\\hline
\multirow{2}{*}{$-1.0$}&$v^s_{T_c},~v^{br}_{T_c}$&$-26.6$, $196.$&$-23.3$, $192.$& $-18.7$, $187.$\\
&$T_c$&$104.$&$102.$& $101.$\\\hline
\multirow{2}{*}{$-1.5$}&$v^s_{T_c},~v^{br}_{T_c}$&$-13.0$, $212.$&$-11.2$, $209.$& $-9.03$, $206.$\\
&$T_c$&$86.3$&$83.9$& $81.9$\\\hline
\multirow{2}{*}{$-2.0$}&$v^s_{T_c},~v^{br}_{T_c}$&$-7.79$, $226.$&$-6.72$, $224.$& $-5.59$ , $221.$\\
&$T_c$&$58.7$&$56.0$& $54.2$\\\hline
\multirow{2}{*}{$-2.5$}&$v^s_{T_c},~v^{br}_{T_c}$&$-$&$-$& $-$\\
&$T_c$&$-$&$-$& $-$\\\hline
\end{tabular}}
\end{center}

\vspace{-0.6cm} 


\begin{center}
{\renewcommand{\arraystretch}{1.2}
\begin{tabular}{ c|c| ccc }
\hline 
{\multirow{2}{*}{$\kappa |v| /m_h^2$}}&&\multicolumn{3}{c}{$|\xi|=0.5\pi$}\\ \cline{3-5}
&& ${y_t}=0.46{y_t^{SM}}$& ${y_t}=0.52{y_t^{SM}}$ &${y_t}=0.57{y_t^{SM}}$  \\\hline\hline

\multirow{2}{*}{$0$}&$v^s_{T_c},~v^{br}_{T_c}$&$-$&$-$&$-$\\
&$T_c$&$-$&$-$&$-$\\\hline
\multirow{2}{*}{$-1$}&$v^s_{T_c},~v^{br}_{T_c}$&$-$&$74.68$, $99.68$& $60.$, $113.94$\\
&$T_c$&$-$&$159.447$& $152.532$\\\hline
\multirow{2}{*}{$-1.5$}&$v^s_{T_c},~v^{br}_{T_c}$&$36.1$, $186.$&$31.3$, $188.$&$16.3$, $190.$\\
&$T_c$&$142.$&$135.$&$128.$\\\hline
\multirow{2}{*}{$-2$}&$v^s_{T_c},~v^{br}_{T_c}$&$6.31$, $218.$&$4.80$, $219.$&$3.66$, $219.$\\
&$T_c$&$115.$&$107.$&$101.$\\\hline
\multirow{2}{*}{$-2.5$}&$v^s_{T_c},~v^{br}_{T_c}$&$-0.666$, $238.$&$-1.33$, $238.$&$-1.83$, $238.$\\
&$T_c$&$77.0$&$67.7$&$58.6$\\\hline
\multirow{2}{*}{$-3$}&$v^s_{T_c},~v^{br}_{T_c}$&$-$&$-$&$-$\\
&$T_c$&$-$&$-$&$-$\\\hline
\end{tabular}}
\end{center}
\vspace{-0.5cm}
 \caption{{\small The thermal expectation values $v^s_{T_c}$, $v^{br}_{T_c}$ and the critical temperature $T_c$ for various values of $\kappa$, $y_t$ and $\xi$ corresponding to the strong first-order phase transition. The dash sign indicates that the transition is a crossover or of the second-order.   Constraints on $y_t$ and $\xi$ obtained in \cite{Kobakhidze:2014gqa} are taken into account. We use $m^{\prime (+)}_t$ from Eq. (\ref{11}) in the first two tables, as no first-order phase transition is found for $m^{\prime (-)}_t$ and positive $v>0$. For $|\xi|=0.5\pi$ in the last table  $m^{\prime (+)}_t=m^{\prime (-)}_t$. Another set of valiable solutions  is obtained by revercing signs of $\kappa$, $v^s_{T_c}$, $v^{br}_{T_c}$ and $m^{\prime (+)}_t\leftrightarrow -m^{\prime (-)}_t$ simultaneously. }}
\label{tab1}
\end{table}


\section{Computing the baryon asymmetry}
Before proceeding to the calculation of the baryon asymmetry in the model with the anomalous Higgs couplings, we recall a qualitative picture of the electroweak baryogenesis within the so-called charge transport mechanism \cite{Nelson:1991ab}. The non-equilibrium electroweak phase transition proceeds through the nucleation and subsequent expansion of bubbles of the broken phase within the surrounding plasma in the symmetric phase. Plasma particles scatter of a bubble wall and these scatterings generate CP (and C) asymmetries in particle number densities in front of the wall, providing the underlying theory is CP non-invariant. The CP asymmetries diffuse into the symmetric phase where the rapid baryon number violating sphaleron transitions \cite{Manton:1983nd, Klinkhamer:1984di} produce more baryons than antibaryons. Finally, the net baryon charge  created outside the bubble wall is swept up by the expanding wall into the broken phase. The sphaleron transitions must be sufficiently suppressed inside the bubble to avoid the wash-out of the generated baryon asymmetry.    

The quantitative description of the above picture is provided by the solutions of the coupled transport equations \cite{Huet:1995sh}. These equations describe the evolution of the net particle number densities, $\delta n_i\equiv n_i-n^c_i\approx \frac{1}{6}g_i\mu_iT^2$,  in a primordial plasma that undergoes the eletroweak phase transition. Here $n_i (n_i^c)$ are particle (antiparticle) number density of the $i$-th particle specie, $g_i$ is the statistical factor (2 for bosons and 3 for fermions in thermal equilibrium), and $\mu_i$ is the chemical potential. The relevant species initially are the left-handed top and bottom quarks with net particle number density $Q\equiv \delta n_{t_L}+\delta n_{b_L}$,  the right-handed top ($T\equiv \delta n_{t_R}$) and the neutral Higgs particle ($H\equiv n_h$), since these are generated through the dominant Higgs-top Yukawa interactions. The individual particle asymmetries can change through the top quark Yukawa interaction at a rate $\Gamma_y$, the top quark mass chirality flip at a rate $\Gamma_m$, the Higgs self interactions at a rate $\Gamma_h$, and weak and the QCD sphaleron interactions at rates $\Gamma_{ws}$ and $\Gamma_{ss}$, respectively. The strong sphaleron interaction rate $\Gamma_{ss}$ is fast enough to maintain chemical equilibrium between left-handed and right-handed quarks, and hence they generate net particle number densities for right-handed bottom $B\equiv \delta n_{b_R}$ as well as for the two light generation quarks: $Q_{a}\equiv \delta n_{u^a_L}+\delta n_{d^a_L}$, $U_a\equiv \delta n_{u^a_R}$ and $D_a\equiv \delta n_{d^a_R}$ ($a=1,2$). These densities are related to the ones of the third generation through Eq. (\ref{appb1}). Therefore, it is suffice to consider the evolution of $Q$, $T$ and $H$, see Eq. (\ref{eqn:diffusion}) . 

The rate $\Gamma_y$ in our scenario gets modified due to the anomalous Higgs-top Yukawa interactions compared to the Standard Model rate  $\Gamma_y^{SM}$: $\Gamma_y=\left(\frac{y_t}{y_t^{SM}}\right)^2\Gamma_y^{SM}\approx (0.2-1.4)\Gamma_y^{SM}$. Even for small $|y_t|$ we found the rate $\Gamma_y$ is still fast enough and we employ the approximation of Ref. \cite{Huet:1995sh} (see also Appendix \ref{B}). 
Ignoring the weak sphaleron rate inside the bubble is  negligible, the net baryon number density in the broken phase is computed to be: 
\begin{eqnarray}
n_B& =& \left(-\frac{3\Gamma_{ws}}{v_w}\right)\left(-\frac{3\overline{D}^{-1}}{64\Gamma_{ss}}\right)\Bigg[D_q\int_{-\infty}^0 dz S^{CPV} _t(z) \nonumber \\
&+& \frac{D_q k_+^2-v_w k_+}{k_+-k_-} \int_{-\infty}^0 dz \int_{-\infty}^{z} dz'\   e^{k_+(z-z')} S^{CPV}_t(z')\nonumber \\
&+ &\frac{D_q k_-^2-v_w k_-}{k_+-k_-}\int_{-\infty}^0 dz\int^{\infty}_{z} dz' \  e^{k_-(z-z')} S^{CPV}_t(z')\Bigg]~, 
\label{15}
\end{eqnarray}
where $z$ is a coordinate normal to the bubble wall in the wall's rest frame, with $z>0$ $(z<0)$ being the broken (symmetric) phase, and $v_w$ is the wall's velocity. $S^{CPV}_t(z)$ is the CP-violating source \cite{Huet:1995mm} related to the Higgs-top anomalous interactions (\ref{5}). It will be discussed in more details below. $\bar D$ in Eq. (\ref{15}) is the effective diffusion constant given defined in terms of quark $D_q$ and Higgs $D_h$ diffusion constants as $\bar D=\frac{1}{16}(9D_q+7D_h)$. The strong sphaleron rate is given by  $\Gamma_{ss}=4.9 \times 10^{-4} T$ \cite{Moore:1997im}, while $k_{\pm}$ is  
\begin{equation}
\begin{aligned}
k_{\pm} &= \frac{v_w}{2\overline{D} } \left (1\mp \sqrt {1+\frac{4\overline{D}{\Gamma}_\pm}{v_w^2}}\right)\\
\label{15b}
\end{aligned}
\end{equation}
where $\Gamma_+$ and $\Gamma_-$ are the effective Higgs decay rate (cf. Eq.~\ref{eqn:diffusion_sol2} in the Appendix) evaluated with  Eq.~\ref{eqn:diffusion_greens} using the appropriate $v_{T_c}^{br}$ and $v_{T_c}^{s}$ respectively. The weak sphaleron rate $\Gamma_{ws}$ is discussed in the next section. 

\subsection{The electroweak sphaleron rate}
\begin{figure}[t]
\begin{center}
\includegraphics[width=0.7\textwidth]{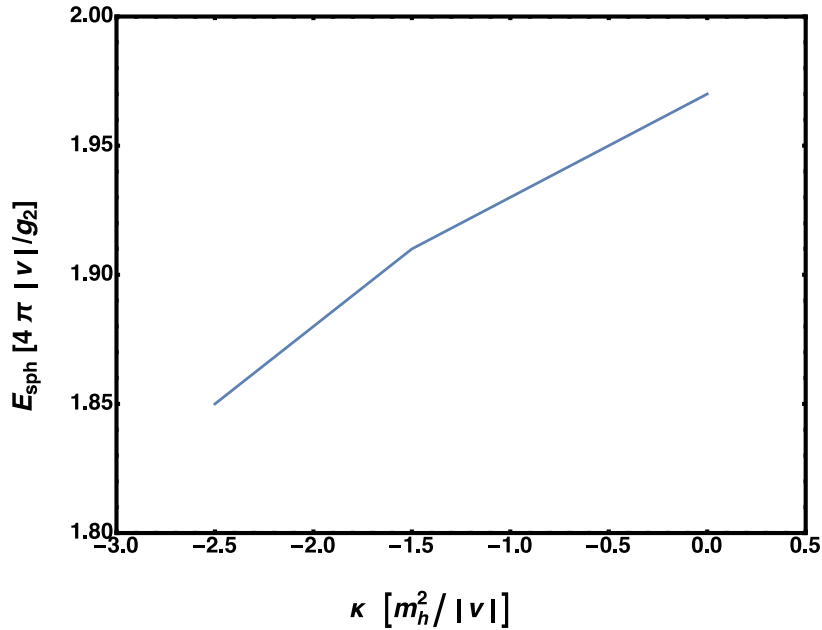}
\end{center}
\caption{{\small The factor $B(g_2, \lambda, \kappa )$ (\ref{17}) as a function of $\kappa$ for fixed $\lambda$ (\ref{12}). The Standard Model value for the factor is $B^{SM}=B(g_2, \lambda, 0 )\approx 1.97$. }}
\label{fig2}
\end{figure}

The finite temperature electroweak sphaleron rate within a thermal volume $V=1/T^3$ in the case of non-zero Higss thermal vacuum expectation value $v_T$ is given by \cite{Arnold:1987mh, Carson:1990jm}: 
\begin{equation} 
\Gamma_{sph}
\approx  \displaystyle\frac{4x\omega_-}{g_2|v_T| T^3}  \left(\frac{\alpha_w T}{4\pi}\right)^4  \mathcal{N}_{tr} (\mathcal{N}\mathcal{V})_{rot} \left(\frac{4\pi |v_T|}{g_2T} \right)^7 \exp \left(-\frac{E_{sph}(T)}{T}\right)
\label{16}
\end{equation}
Here $\alpha_w=g^2_2/4\pi \approx 1/29.5$ is the weak isospin fine structure constant; $\omega_{-}\approx g_2|v|$ is a dynamical pre-factor 
which is related to the absolute value of the negative eigenvalue of the fluctuation operator around the sphaleron solution; $\mathcal{N}_{tr}\approx 26$ and $(\mathcal{N}\mathcal{V})_{rot} \approx 5.3 \times 10^3$ \cite{Arnold:1987mh} represent  normalisation factors related to the translational and rotational zero-modes and $x\approx 0.03$ contains the contributions of the positive modes of the fluctuation operator \cite{Carson:1990jm}. The numerical values for the above factors are estimated within the SM for $\sqrt{\lambda}= g_2$ and, to a reasonable  accuracy, serve our case as well, especially for not too large anomalous cubic coupling\footnote{According to Eq. (\ref{12}), the Higgs quartic coupling $\lambda$ in our case is larger than the Standard Model one, $\lambda^{SM}\approx 0.129$. Hence, $\sqrt{\lambda}/g_2\sim 0.7-1$ for $|\kappa|\sim 30-160$ GeV}. This readily follow from the fact that sphaleron solution in our case does not differ much from the one in the SM, as it is explicitly demonstrated in Appendix \ref{C}. Consequently, the zero temperature sphaleron energy 
\begin{equation}
E_{sph}(T=0)=\frac{4\pi |v|}{g_2}B(g_2, \lambda, \kappa )~,
\label{17} 
\end{equation}
is close to the corresponding value in the Standard Model [$B^{SM}=B(g_2, \lambda, 0 )\approx 1.97$] for a wide range of $\kappa$ parameter. See Figure \ref{fig2} for $\kappa$ dependence of $B(g_2, \lambda, \kappa )$ with $\lambda$ fixed as in Eq. (\ref{12}). 

The sphaleron energy at finite temperature is computed by scaling the zero temperature energy (\ref{17}):
\begin{equation}
E_{sph}(T)=E_{sph}\frac{|v_T|}{|v|}\sim 37 |v_T|~,
\label{18}
\end{equation}
for $-\kappa \in [6, 160]~{\rm GeV}$. Recall that Eq. (\ref{16}) with estimation (\ref{18}) is applicable for both broken and symmetric phase. In the symmetric phase we would like to avoid suppression  in (\ref{16}) during the phase transition.  
By demanding that the rate (\ref{16}) is comparable or larger than the Standard Model rate $\Gamma^{SM}\sim \alpha_w^4 T$ at a critical temperature $T=T_c$, we obtain:
\begin{equation}
\left(\frac{|v_{T_c}^s|}{T_c}\right)\exp \left(-\frac{37}{7}\frac{|v_{T_s}^s|}{T_c}\right) \gtrsim 0.055  \left(\frac{|v^{s}_{T_c}|}{|v|}\right)^{1/7}
\label{19}
\end{equation}  
To avoid a large exponential suppression of the left-hand side of this equation we must require $v_{T_s}^s|/T_c \lesssim 1$. However, for  
 $v_{T_s}^s|/T_c \lesssim 0.26$ the the non-exponential factor becomes even smaller. Therefore, we find a finite range of $|v_{T_s}^s|/T_c$ for which the sphaleron transition is unsuppressed in the symmetric phase:
 \begin{equation}
0.055  \left(\frac{|v^{s}_{T_c}|}{|v|}\right)^{1/7}\lesssim \frac{|v_{T_s}^s|}{T_c} \lesssim 1~.
 \label{20}
 \end{equation}
 In the broken phase we require $|v^{br}_{T_c}|/T_c\gtrsim 1$, so the shpalerons are ineffective inside the bubble. 
 
 \subsection{CP-violation via the anomalous Higgs-top couplings}
Next we discuss the CP-violating source $S^{CPV}_t(z)$ entering the formular for the baryon number density in Eq. (\ref{15}). 
In our model it originates from the CP-violating anomalous Higgs-top Yukawa interactions (\ref{5}) and takes the form \cite{Huet:1995mm}:
\begin{equation}
S^{CPV}_{t} (z)\approx \frac{3}{2\pi^2} T\gamma_w v_w m_t^2(z)  \partial_z\theta_t(z)~,
\label{21}
\end{equation}
where $\gamma_w=1/\sqrt{1-v_w^2}$; $m_t(z)$ and $\theta(z)$ are the modulus and phase of the complex top quark mass, $M_t(z) = m_t(z) e^{i\theta_t(z)}$, 
\begin{eqnarray}
m_t(z)&=&\sqrt{\left(m'_t+\frac{y_t}{\sqrt{2}} h(z) \cos \xi \right)^2 +\left( \frac{y_t}{\sqrt{2}} h(z) \sin \xi \right )^2}\\
\tan\theta_t(z)&=&\frac{{y_t} h(z) \sin \xi}{\sqrt{2}m'_t+{y_t}h(z) \cos \xi}
\label{22}
\end{eqnarray}
 in the background of the bubble wall $\rho(z)$. We note that the CP-violation is entirely defined by a phase $\xi$ of the anomalous interactions (\ref{5}), $S^{CPV}_{t} \to 0$ as $\xi\to 0.$ 
 
 Ignoring the wall curvature, the bubble wall configuration $\rho(z)$ at a critical temperature  has the following simple form:    
\begin{equation}
\rho(z) =v^{br}_{T_c} + \frac{v^{s}_{T_c}-v^{br}_{T_c}}{2} \left[1+\tanh\left(\frac{z}{L_w}\right)\right]~. 
\label{23}
\end{equation}
This configuration approaches the broken vacuum $v^{br}_{T_c}$ as $z\to -\infty$ (inside the bubble) and the symmetric vacuum $v^{s}_{T_c}$ as $z\to +\infty$ (outside bubble), as it should be. The wall width $L_w$  can be analytically computed when the effective potential is approximated by Eq. (\ref{14}):
\begin{equation}
L_w=\frac{3\sqrt{\lambda}}{|2\kappa+6\lambda v^s_{T_c}|}\approx \frac{3m_h}{|\kappa v|}\sqrt{1+\frac{|v\kappa|}{m_h^2}}~,
\label{24}
\end{equation}
 where in the last step we assume $v^s_{T_c}<<|v|$. We observe that the width of the wall is essentially defined through the anomalous cubic coupling $\kappa$ -- larger is $\kappa$, thinner is the wall. More precise definition of the wall width requires numerical calculations. Also, the determination of the wall velocity requires more careful study of the dynamics of the wall. The typical calculations in various models contain large theoretical uncertainties (see, e.g., \cite{Anderson:1991zb}-\cite{Megevand:2009gh}). We do not attempt to perform such calculations in this paper.  Instead, we keep $L_w$ and $v_w$ as parameters and vary them in the expected ranges $L_{w}\in [3/T_c, 16/T_c]$ and     
$v_w\in [10^{-3}, 1/\sqrt{3}]$. The lower bound on $v_w$ is to ensure that the expanding wall creates non-equilibrium, while the upper bound corresponds to the sound speed in relativistic plasma. The wall must move subsonic in order the CP asymmetric particle number densities, created in front of the wall, to diffuse efficiently into the symmetric phase.

\subsection{The baryon asymmetry}
\begin{figure}[t]
\includegraphics[width=0.5\textwidth]{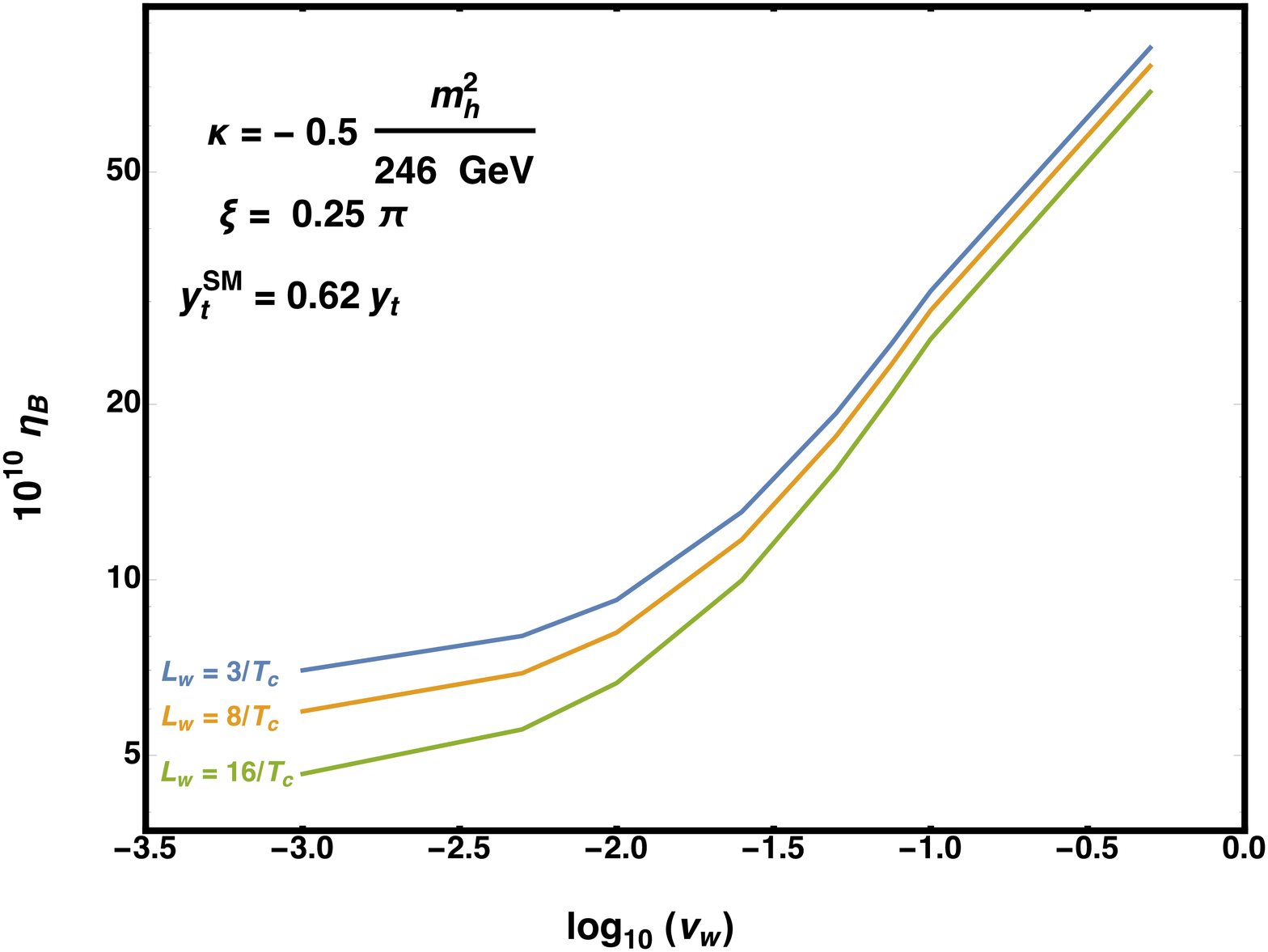} 
\includegraphics[width=0.5\textwidth]{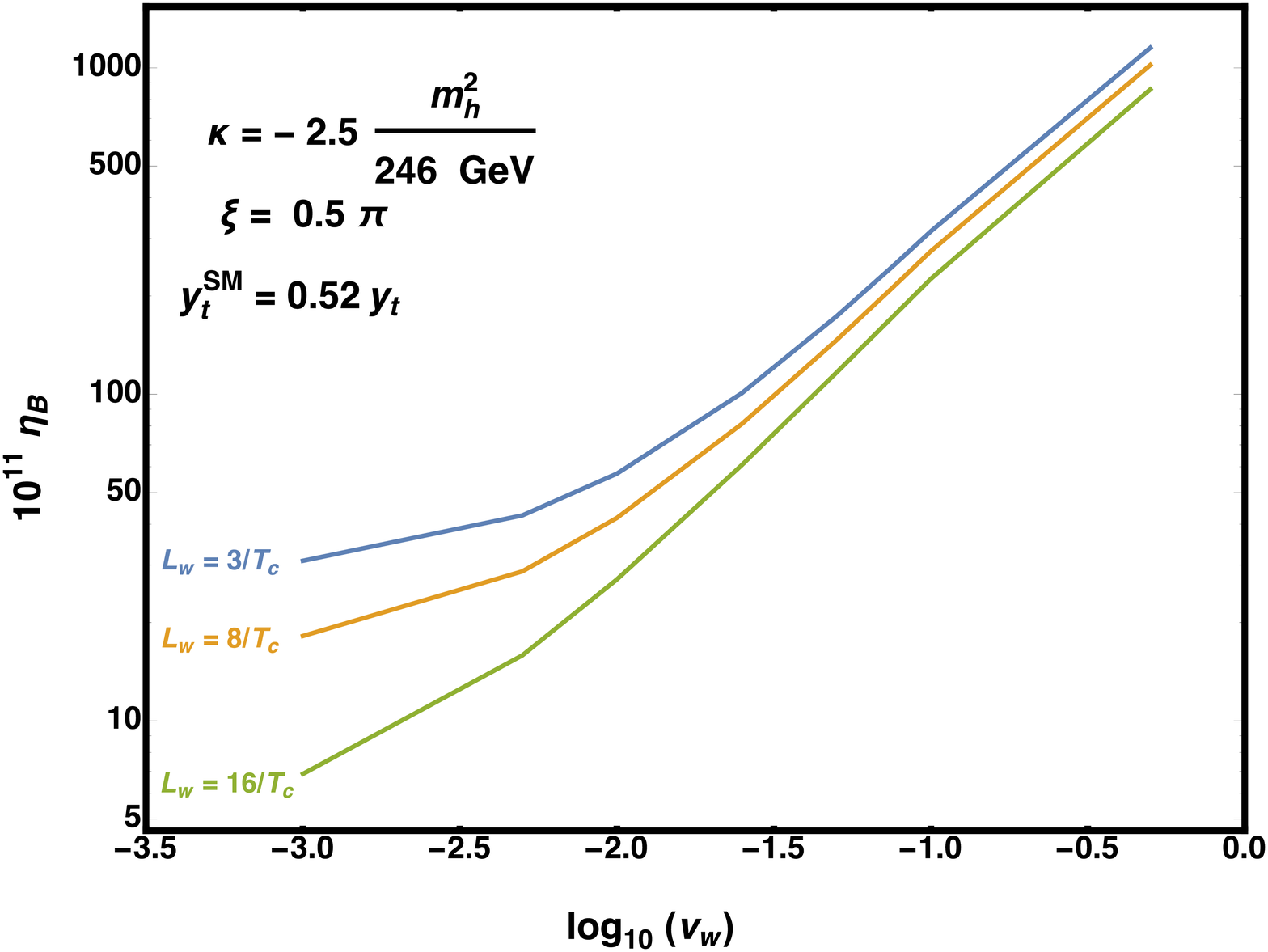}
\caption{{\small The dependence of the baryon asymmetry parameter $\eta_B$ on the width of the bubble wall $L_w$ and its velocity $v_w$. The wall velocity is bounded $0.001\lesssim v_w \lesssim 1/\sqrt{3}$.}}
\label{fig3}
\end{figure}

We have all the ingredients now to compute the asymmetry parameter, $\eta_B=n_B/s$, where $n_B$ is computed in Eq. (\ref{15}) and $s=2\pi^2 g_* T^3 /45$ is the entropy density with $g_*\sim 100$ counting the effective number of relativistic degrees of freedom in equilibrium at temperature $T$. The results of numerical calculations are presented in Table \ref{tab2} and Figures \ref{fig3}. 

We observe that a significant asymmetry can be produced assuming the Higgs anomalous couplings $\kappa$, $y_t$ and phase $\xi$. 
The asymmetry increases with the increase of the anomalous couplings $\kappa$ and $y_t$, the later being subject to the constraints obtained in \cite{Kobakhidze:2014gqa}. However, for large $\kappa$ either the first-order phase transition fails, or Eq. (\ref{20}) is not satisfied and the sphaleron rate in the symmetric phase reduces significantly. Thus, there is only a finite range of $\kappa$ for which a significant baryon asymmetry can be produced. There are some large theoretical uncertainties in our calculations and, therefore, we do not attempt here at more thorough scanning of the parameter area in order to determine the range of parameters which reproduces the observed asymmetry in Eq. (\ref{1}). For example, the bubble wall width $L_w$ and its velocity $v_w$ are treated as free parameters, rather than being defined by studying the dynamics of the wall. In Figure \ref{fig3} we have plotted the dependence of the baryon asymmetry $\eta_B$ on $v_w$ and $L_w$. A relatively large variation in the asymmetry parameter is observed for small $v_w$. With the increase of $v_w$ the asymmetry becomes less dependent on $L_w$.      

\begin{table}[t]
\fontsize{8pt}{9.6pt}\selectfont
\begin{center}
{\renewcommand{\arraystretch}{1.5}
\begin{tabular}{ c|c|c| c|c }
\hline 
&{\multirow{2}{*}{$\kappa |v| /m_h^2$}}&\multicolumn{3}{c}{$\xi=0.25\pi$}\\ \cline{3-5}
&& ${y_t}=0.62{y_t^{SM}}$& ${y_t}=0.69{y_t^{SM}}$ &${y_t}=0.76{y_t^{SM}}$  \\\hline\hline
{\multirow{4}{*}{$\eta_B=n_B/s $ }}&$-0.5$&$9.28\times 10^{-10}$&$3.41\times 10^{-9}$&$2.41\times 10^{-8}$\\\cline{2-5}
&$-2.0$&$2.02\times 10^{-5}$&$1.92\times 10^{-5}$&$1.38\times 10^{-5}$\\\cline{2-5}
&$0.5$&$8.31\times 10^{-10}$&$3.98\times 10^{-9}$&$3.35\times 10^{-8}$\\\cline{2-5}
&$2.0$&$6.18\times 10^{-6}$&$1.00\times 10^{-5}$&$1.01\times 10^{-5}$\\\hline
\end{tabular}}
\end{center}
\vspace{-0.6cm}

\begin{center}
{\renewcommand{\arraystretch}{1.5}
\begin{tabular}{ c|c|c| c|cc }
\hline 
&{\multirow{2}{*}{$\kappa |v| /m_h^2$}}&\multicolumn{3}{c}{$\xi=0.5\pi$}\\ \cline{3-5}
&& ${y_t}=0.46{y_t^{SM}}$& ${y_t}=0.52{y_t^{SM}}$ &${y_t}=0.57{y_t^{SM}}$  \\\hline\hline
{\multirow{3}{*}{$\eta_B=n_B/s $ }}&$-1.5$&$1.14\times 10^{-6}$&$1.72\times 10^{-6}$&$1.59\times 10^{-6}$\\\cline{2-5}
&$-2.0$&$8.02\times 10^{-8}$&$3.48\times 10^{-8}$ &$1.24\times 10^{-8}$\\\cline{2-5}
&$-2.5$&$2.10\times 10^{-12}$&$5.71\times 10^{-10}$&$1.21\times 10^{-8}$\\\hline
\end{tabular}}
\end{center}
\vspace{-0.5cm}
\caption{{\small Representative numerical values of baryon asymmetry parameter $\eta_B$ computed with fixed  $v_w=0.01$ and $L_w = 3/T_c$ and for (i) $\xi=0.25\pi$, various values of $y_t$ subject to constraints obtained in \cite{Kobakhidze:2014gqa} and  $\kappa|v|  /m_h^2= -0.5, -2$ (with $v=+246$ GeV and ${m}^{\prime (+)}_t$) and $\kappa|v|  /m_h^2  = 0.5, 2$ (with $v=-246$ GeV and ${m}^{\prime (-)}_t$);  (ii) $\xi=0.5\pi$, various values of $y_t$ subject to constraints from \cite{Kobakhidze:2014gqa} and $\kappa |v| /m_h^2 = -0.5, -2, -2.5$ (with $v=+246$ GeV and ${m}^{\prime (+)}_t ={m}^{\prime (-)}_t$). In this case the same results are obtained  by reversing signs of $\kappa$ and $v$ simultaneously. }}
\label{tab2}
\end{table}

\section{Conclusion and outlook}
In this paper we have studied the electroweak phase transition and computed the baryon asymmetry in the Standard Model with the anomalous Higgs cubic self interactions with strength $\kappa$ and CP-violating Higgs-top Yukawa interactions defined by modulus $y_t$ and phase $\xi$. The model is a restricted case of more general Standard Model with the electroweak symmetry being non-linearly realised. As an effective theory, the model is valid up to relatively high scales, $\sim 10$ TeV. In the worth case scenario, the underlying physics may even escape the detection at the LHC. In addition, the model is relatively less constrained from the electroweak precision and electric dipole moment measurements and the current LHC data. 

We have found the strong first-order phase transition is realised for a wide range of the anomalous couplings (see Table {\ref{tab1}}). The notable point of our scenario is that the Higgs expectation value $v^s_{T_c}$ does not vanish in the symmetric phase, see Eq. (\ref{14a}) and Figure \ref{fig1}. This has important implications for baryogenesis as one must ensure sufficiently fast baryon number violating sphaleron transitions take place in the symmetric phase. We found a finite range (\ref{20}) of $v^s_{T_c}$ for which the sphaleron transitions are not suppressed. The necessary for baryogenesis CP violation is governed by the phase $\xi$. We then computed the baryon asymmetry parameter $\eta_B$ and observed that the measured asymmetry (\ref{1}) can be comfortably accommodated within the allowed range of parameters of the anomalous interactions (see Table \ref{tab2} and Figure \ref{fig3}).

The critical experimental test of the proposed scenario comes with the measurements of the Higgs-top Yukawa and Higgs cubic couplings. The Higg-top Yukawa coupling can be measured at LHC in processes of the Higgs associated production with single top (antitop)  quark ($htj$) or piar of top-antitop quarks ($h\bar tt$). In the Standard Model the $h\bar tt$ production cross section is an order of magnitude larger than the $htj$ production cross section. However, in the presence of a significant pseudoscalar couplings, as it is required by the baryogenesis scenario, $h\bar tt$ cross section decreases, while  $htj$ cross section increases, such that  for $|\tan \xi | \gtrsim 1$ the $htj$ cross section becomes larger \cite{Ellis:2013yxa}. Thus, accurate measurements of the total $h\bar tt$ and $htj$ cross sections would provide an important indirect evidence on the validity of our scenario. Another observable which carries the information on the anomalous Higgs-top Yukawa couplings is the three-body invariant mass distribution for $h\bar tt$ and $htj$. As $|\tan \xi|$ increases,  the $h\bar tt$ invariant mass distribution becomes less peaked at small masses, while the $htj$ invariant mass distribution exhibits the opposite behaviour. The CP-violating nature of the anomalous Higgs-top Yukawa couplings can also be probed by measuring the angular distributions of leptons resulting from polarised top quark decays in $htj$ channel \cite{Kobakhidze:2014gqa, Yue:2014tya}.  Although promising, the measurements of the anomalous Higgs-top Yukawa couplings are rather challenging at LHC due to the large backgrounds. So together with further improvements in experimental techniques and theoretical calculations (see, e.g., \cite{Demartin:2014fia, Maltoni:2015ena} for recent studies), one requires high luminosity to obtain sizeable sensitivity to the anomalous Higgs-top Yukawa couplings. Other recent studies on the anomalous Higg-top interactions can be found in \cite{Degrande:2012gr}-\cite{Moretti:2015vaa}. 

The measurement of the Higgs cubic coupling at the LHC is even more challenging. The best way to probe the cubic Higgs coupling can be is  through the radiative process of double Higgs production (for recent works see \cite{Baur:2003gp}-\cite{Lu:2015jza}).  At 1-loop level there are two diagrams responsible for this process: triangle diagram which involves Higgs-top Yukawa and Higgs cubic couplings and the box diagrams which involve only Higgs-top Yukawa coupling. Thus, in order to probe the effects of the Higgs cubic coupling, one must to disentangle the contributions from the 
triangle and box diagrams. 

With the anomalous couplings favourable by our baryogenesis scenario, the cross section  for di-Higgs production can be enhanced by $\sim 3-10$ times  relative to the SM value \cite{Liu:2014rba, Lu:2015jza}, especially for the negative cubic coupling\footnote{For positive $\kappa^{SM} \lesssim \kappa \lesssim 5\kappa^{SM}$ [$\kappa^{SM}=\frac{3m_h^2}{|v|}$ is the SM value for the cubic coupling] there is a partial cancellation between triangle and box diagrams \cite{Frederix:2014hta}.}. The recent studies show that the anomalous Higgs coupling $\kappa$ can be determined at 14 TeV high-luminosity (3000 fb$^{-1}$) LHC with 25-50\% accuracy at most \cite{Lu:2015jza}, due to the entanglement of $\kappa$ and $y_t$ parameters in the di-Higgs production cross section. Hence, to probe the cosmological eletroweak phase transition and the related mechanism for baryogenesis, we ultimately need to resort to new colliders, such as the planned International Linear Collider or/and more powerful 100 TeV hadron collider.           

\paragraph{Acknowledgement.} We would like to thank Masaki Asano, David Curtin, Manuel Drees, Mark Hindmarsh, Tao Liu and Nicholas Manton for useful discussions. This work was partially supported by the Australian Research Council. AK was also supported in part by the Rustaveli National Science Foundation under the project No. DI/12/6-200/13.

\appendix
\section{Finite temperature effective potential}
\label{A}
Finite temperature 1-loop corrections to the tree-level potential are encoded in $\delta V^{(1)}_T$ in Eq. (\ref{14}). This term comprises of 1-loop zero temperature Coleman-Weinberg potential \cite{Coleman:1973jx}, $\delta V_{CW}$, and the one 1-loop thermal potential $\delta V_T$ \cite{Dolan:1973qd}, $\delta V^{(1)}_T=\delta V_{CW}+\delta V_T$. The 1-loop Coleman-Weinberg potential computed using the $\overline{\rm DR}$ subtraction scheme, looks as:       
\begin{eqnarray}\label{eqn:one_loop}
\delta V_{CW}(\rho)&=&\sum_{i=\{\rho, W, Z, t\}}\frac{n_i m^4_i}{64 \pi^2} \left(\ln\left(  \frac{ m^2_i(\rho)}  {\mu_R^2}\right)-\frac{3}{2}\right)~, 
  \end{eqnarray}
 where $i$ runs over the the fields which give the dominant contribution to the potential, i.e., $\{\rho, W, Z, t\}$\footnote{We work in the unitary gauge by setting the Goldstone fields $\pi_i=0$ and do not worry about the subtleties related with the gauge dependence of the effective potential. Anyway, the contribution of Goldstone fields seems to be numerically insignificant \cite{Elias-Miro:2014pca, Martin:2014bca,Pilaftsis:2015bbs}.}, and their respective degrees of freedom are: 
    \begin{eqnarray}\label{eqn:dof}
   n_{\{\rho,W,Z,t\}}=\{1,6,3,-12\}~.
     \end{eqnarray}
 The renormalisation scale $\mu_R$ is taken to be $|v|$ and field-dependent masses $m^2_i(\rho)$ are:
 \begin{eqnarray}
 m_W^2(\rho)&=&g_2^2\rho^2/4~,~~m_Z^2(\rho)=(g_2^2+g_1^2)\rho^2/4~, \label{effmass}\\
 m_h^2(\rho)&=&-\mu^2+2\kappa \rho +3\lambda \rho^2~, \label{effmassh} \\
 m_t^2(\rho)&=&\left(m'_t+\frac{y_t\rho\cos\xi}{\sqrt{2}}\right)^2+\frac{y_t^2}{2}\rho^2\sin^2\xi~.
 \label{effmasst}
 \end{eqnarray}   
Note that, due to the anomalous interactions, the Higgs and top-quark tree-level masses are no longer proportional to the Higgs expectation value $v$. 

The 1-loop finite temperature potential takes the form:
 \begin{eqnarray}
\delta V_T(\rho)= \frac{T^4}{2\pi^2} \sum_{i=h,W,Z,t} n_i J_i\left[\frac{m_i^2(\rho)}{T^2}\right]~,
\label{effthermal}
\end{eqnarray}     
where, 
  \begin{eqnarray}\label{ref:temperature_func}
J_i(m_i^2(\rho)/T^2) &=&\int^\infty_0 dp \ p^2 \ln \left[1-(-1)^{2s_i}e^{-\sqrt{p^2+m_i^2(\rho)}/T}\right]  
\end{eqnarray}
with $s_i$ denoting the spin of the $i$-particle. In the high-temperature limit $T\gg m_i(\phi)$ we obtain\footnote{For $T\gg m_i(\phi)$, the bosonic logarithm in \ref{ref:temperature_func} diverges in the infrared $p\to 0$. This infrared divergences are dealt by resuming the multi-loop bosonic contributions which result in the so-called ring-terms in the effective potential. We will not consider them here.}:
\begin{equation}
    J_i(x)= \begin{cases}
      -\frac{\pi^4}{45} +\frac{\pi^2}{12} x-\frac{\pi}{6} x^{3/2}+ \mathcal{O}\left(x^2\right), &\text{for bosons}\\  
        \frac{7\pi^4}{360} -\frac{\pi^2}{24} x +\mathcal{O}\left(x^6\right), &  \text{for fermions}
\end{cases}
\end{equation}
These expansions have been used in deriving Eq. (\ref{14}). In particular, the linear term $\sim T^2\rho$ originates from the $\sim x=m_i (\rho)^2/T^2$ terms in the above expansions and the Higgs and top quark field-dependent masses, Eqs. (\ref{effmassh}, \ref{effmasst}).  
Note that this linear term is different from the spurious $\sim T^3\rho$ term discussed in \cite{Dine:1992wr}.

\section{Transport equations}
\label{B}
The most general set of transport equations are rather complicated to deal with. The following simplifications can be made based on the physics motivated approximations. The strong sphaleron transitions tend to populate plasma with right-handed bottom quarks and left- and right-handed quarks of the first two generations. The corresponding net particle densities are related with those of left-handed top and bottom and right-handed top quarks as:     
\begin{equation}
Q+T = -B  = -U_a = -D_a = Q_a/2~.
\label{appb1}
\end{equation}
As a result it is sufficient to consider the transport equations for $Q$, $T$ and $H$ only \cite{Huet:1995sh}:  
\begin{align}
\label{eqn:diffusion}
D_q Q''- v_w Q' -\Gamma_y\left(\frac{Q}{g_Q}-\frac{H}{g_H}-\frac{T}{g_T}\right)-\Gamma_m\left(\frac{Q}{g_Q}-\frac{T}{g_T}\right)\nonumber\\
-6\Gamma_{ss}\left(\frac{2Q}{g_Q}-\frac{T}{g_T}+\frac{9(Q+T)}{k_B}\right)+S^{CPV}_{t} =0\nonumber\\
D_q T''- v_w T' -\Gamma_y\left(-\frac{Q}{g_Q}+\frac{H}{g_H}+\frac{T}{g_T}\right)-\Gamma_m\left(-\frac{Q}{g_Q}+\frac{T}{g_T}\right)\nonumber \\
-3\Gamma_{ss}\left(-\frac{2Q}{g_Q}+\frac{T}{g_T}-\frac{9(Q+T)}{k_B}\right)-S^{CPV}_{t} =0\nonumber\\
D_h H''- v_w H' -\Gamma_y\left(-\frac{Q}{g_Q}+\frac{H}{g_H}+\frac{T}{g_T}\right)-\Gamma_{h}\left(\frac{H}{g_H}\right) =0
\end{align}
Here $\prime$ denotes differentiation wrt coordinate $z$, $\partial_z = v_w\partial_t$ and 
\begin{align}
\label{eqn:stat_fact_sm}
g_Q=2g_T=2g_B = 3g_H=6~. 
\end{align}
These equations are further simplified in the limit of large $\Gamma_{ss}$, $\Gamma_y \to \infty$. In fact, the first two equations are reduced to algebraic expressions, which allows us to express $Q$ and $T$ in terms of $H$: 
\begin{align}\label{eqn:diffusion_sol}
{Q} =  \frac{24}{37}H+\mathcal{O}\left(\frac{1}{\Gamma_{ss}},\frac{1}{\Gamma_y}\right)~,~~{T} =  -\frac{30}{37}H +\mathcal{O}\left(\frac{1}{\Gamma_{ss}},\frac{1}{\Gamma_y}\right)~.
\end{align}
Plugging this into the third equation of (\ref{eqn:diffusion}) we obtain the equation solely for $H$:
\begin{align}\label{eqn:diffusion_sol1}
-v_w H' + \overline{D} H'' - \overline{\Gamma} H + \overline{S}^{CPV}_t =0~,
\end{align}
where:
\begin{equation}
\begin{aligned}\label{eqn:diffusion_sol2} \color{red}
 &\overline{D} = \frac{9D_q + 7D_h}{16}~,\\
  &\overline{\Gamma}= \frac{7(\Gamma_m +\Gamma_h) }{32 }~,\\
   &\overline{S}^{CPV}_t=\frac{7{S}^{CPV}_t }{16 }~,
   \end{aligned}
\end{equation}
and 
\begin{eqnarray}\label{eqn:diffusion_greens}
 &D_h =\frac{20}{T}~,~~ D_q = \frac{6}{T}\\
 &\Gamma_m = \frac{m_t^2(\rho(z),T)}{63T}~,~~\Gamma_h = \frac{m_W^2(\rho(z),T)}{50T} 
\end{eqnarray}
The solution to Eq. (\ref{eqn:diffusion_sol1}) can be found by using the Green's function method:
\begin{align}\label{eqn:diffusion_greens1}
 H(z) = \frac{\overline{D}^{-1}}{k_+-k_-} \left[  \int^{z}_{-\infty} dz_0 \ e^{k_+(z-z_0)} \overline{S}^{CPV}_t(z_0)
 + \int^{\infty} _{z}dz_0 \ e^{k_-(z-z_0)} \overline{S}^{CPV}_t(z_0)
 \right]
\end{align}
Using Eqs. (\ref{appb1}) and (\ref{eqn:diffusion_sol}) we find that the net left-handed number density $n_L(z) =Q(z) +Q_1(z) +Q_2(z) = 5Q(z) +4T(z)\approx \mathcal{O}\left(\frac{1}{\Gamma_{ss}},\frac{1}{\Gamma_y}\right)$. Evaluating the leading $\mathcal{O}\left(\frac{1}{\Gamma_{ss}}\right)$ (requires assumption that $\Gamma_y \gg \Gamma_{ss}$ ) contribution we obtain:
\begin{align}\label{eqn:diffusion1}
n_L(z)=-\frac{3}{28} \left(\frac{D_q H''(z) -v_wH'(z)}{\Gamma_{ss}}\right)
\end{align}
The net left-handed quark number density is converted into the net baryon density $n_B$ via the weak sphaleron processes. This process is described by the equation:
\begin{align}\label{eqn:diffusion_baryon_density}
D_q n''_B(z) - v _w n'_B(z)  -3 \Gamma_{ws}(z)  n_L (z)=0~, 
\end{align}
solution to which is given in the main text in Eq. (\ref{15}).

\section{The electroweak sphaleron} 
\label{C}
In this Appendix we demonstrate that the standard electroweak sphaleron solution \cite{Manton:1983nd, Klinkhamer:1984di} does not significantly modifies in our model. Since we are assuming the Higgs-gauge interactions are the same as in the SM, the only modification to the standard analysis 
comes from the cubic Higgs self interaction term in the potential (\ref{6}). Setting the fermion fields and time component of gauge fields to zero and working in the $SU(2)$ limit, i.e., $\theta_W\to 0$, the electroweak sphaleron energy of at zero temperature is given by \cite{Manton:1983nd, Klinkhamer:1984di}:
\begin{align}
\label{eqn:sph_energy}
E_{sph}= \int d^3x \left(\frac{1}{4} {W}^a_{ij}{W}^{a}_{ij}+ \frac{1}{2}\partial_i\rho\partial_i\rho+ \frac{\rho^2}{2}(D_i\Sigma )^\dagger(D^i \Sigma ) +V(\rho)\right)~,
\end{align}
where 
\begin{align}\label{eqn:b}
{W}^a_{\mu\nu}: &= \partial_\mu W^a_\nu -\partial_\nu W^a_\mu - i g_2 \epsilon ^{abc} W^b_\mu W^c_\nu,\\
D_\mu&:=\partial_\mu-\frac{i}{2}g_2\sigma^a W^a_\mu 
\end{align}
are the $SU(2)$ field strengths for $W^a_\mu$ gauge fields and  
and  $V(\rho)$  is given by (\ref{6}) with an additional constant that adjust such that the integrand approaches zero asymptotically. The static sphaleron solution can be search for by taking an $O(3)-$symmetric \textit{Ansatz} of the following form:
\begin{align}\label{eqn:sph_param}
\frac{i}{2}g_2\sigma^aW^a_i dx^i &= f_W(\zeta) d U^\infty (U^\infty)^{-1}\\
 \phi&= \frac{v}{\sqrt{2}} f_h(\zeta) U^\infty \begin{pmatrix}
     0     \\     1  
\end{pmatrix},
\end{align}
where $\zeta$ is a dimensionless radial distance $\zeta:=g_2|v|r$ and 
\begin{align}\label{eqn:U_mat}
U^\infty:=\frac{1}{r} \begin{pmatrix}
  z&x+iy \\
  -x+iy &z  
\end{pmatrix}.
\end{align}
The functions $f_W,~f_h$ must satisfy the following equations of motion:
\begin{align}\label{field1}
\zeta^2 \frac{d^2 f_W}{d\zeta^2}&=  2 f_W(1-f_W)(1-2f_W) -\frac{\zeta^2}{4} f_h^2(1-f_W)\\
 \label{field2} \frac{d}{d\zeta}\left(\zeta^2 \frac{d  f_h}{d\zeta}\right)&=  2 f_h(1-f_W)^2+\frac{\zeta^2}{g_2^2 }\left({\lambda} f_h^3+\frac{\kappa}{v} f_h^2- \frac{\mu^2}{v^2} f_h\right)~, 
\end{align}
with boundary conditions 
\begin{align}\label{field_bc}
 \lim_{\zeta\rightarrow 0} (f_W, f_h)=  0 \\ 
\lim_{\zeta\rightarrow \infty} (f_W, f_h)=  1 
\end{align}
that guarantee the finiteness of the sphaleron energy. In terms of $f_W,~f_h$ functions the sphaleron energy takes the form:
\begin{align}
\label{energy}
\nonumber E_{sph}=& \frac{4\pi |v|}{g_2} \int^\infty_0 d\zeta \left[4\left( \frac{df_W}{d\zeta}\right) ^2 +\frac{8}{\zeta^2} [f_W(1-f_W)] ^2+\frac{1}{2}\zeta^2\left( \frac{df_h}{d\zeta}\right) ^2 +[f_h(1-f_W)]^2 \right.\\
 &\left.+\frac{\zeta^2}{g_2^2 v^4}\left(\frac{\lambda}{4} (v f_h)^4+\frac{\kappa}{3} (v f_h)^3- \frac{\mu^2}{2}(v f_h)^2+c\right) \right].
\end{align}

The numerical solutions to Eqs. (\ref{field1}) and (\ref{field2}) are presented in Figure 4 for $\kappa=-0.5m_h^2/|v|$. We have also plotted the SM  solutions ($\kappa=0$) with dashed curves. These two sets of solutions are very close to each other and curves in Figure 4 essentially overlap.  
Having found numerical solutions for the electroweak sphaleron, we have then computed the sphaleron energy (\ref{energy}) for various $\kappa$. The results are shown on Figure \ref{fig2}. 


\begin{figure}[hpt]
\label{fig4}
\begin{center}
\includegraphics[width=.7\textwidth]{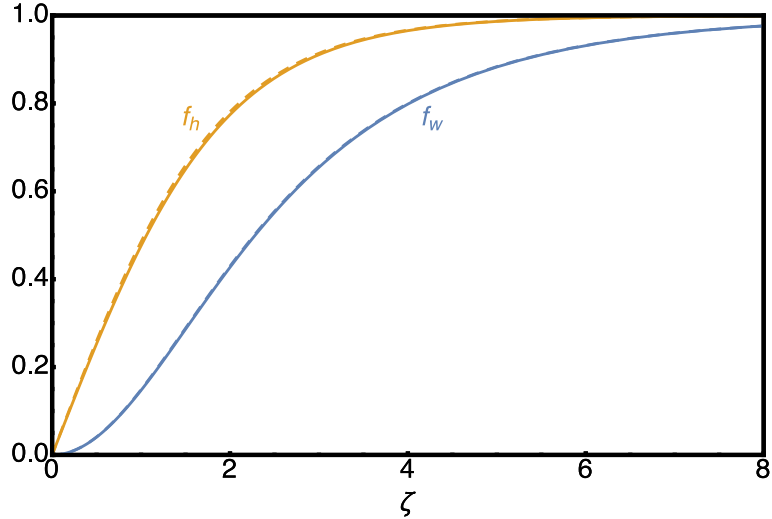}
\end{center}
\caption{\small{ The sphaleron functions $f_W,~f_h$ satisfying  (\ref{field1}) and (\ref{field2}) for $\kappa=-0.5\frac{m_h^2}{|v|}\approx -32$ GeV. The SM solutions ($\kappa=0$) are also plotted with dashed curves. }}
\end{figure}



\end{document}